\documentclass [12pt]{article}
\usepackage {graphicx}
\usepackage {longtable}

\begin{document}
\begin{center}
\textbf{\large Evaluation of Neutrinos Mass Based on ENU Model}
\end{center}

\bigskip

\begin{center}
Miroslav S\'{u}ken\'{\i}k and Jozef \v{S}ima

CHTF STU, Radlinsk\'{e}ho 9, 812 37 Bratislava, Slovakia

\underline {sima@chtf.stuba.sk}
\end{center}

\bigskip

\textbf{Abstract}. Based on principles of the Expansive Nondecelerative 
Universe model that enables to quantify and localize the gravitational 
energy density, and stemming from the ``see-saw'' mechanism, the mass of 
electron, muon and tau neutrinos are determined in an independent way.

\bigskip

\section*{Introduction}
\label{sec:introduction}

\bigskip

For a long time, the neutrinos rest mass had been supposed to be of zero 
value [1]. The opinion has gradually changed, the experimentally observed 
defficiency of solar neutrinos acting as a driving force of the change. 
Mutual oscillations of muon $\nu _{\mu}  $, tau $\nu _{\tau}  $ and electron 
$\nu _{e} $ neutrinos was proposed as a key mode of the defficiency 
interpretation which, in turn, required a nonzero rest mass of the 
neutrinos. In our previous paper [2] we have rationalized this defficiency 
in a different way. There are several approaches to neutrinos masses 
evaluation. Calculations related to some experiments, e.g. tritium decay 
have led to an upper limit of the electron neutrino mass. Based on the 
symmetry between bosons and fermions, and on the relation between the mass 
of neutrinos and corresponding fermions, we have performed calculation of 
the mass of all three neutrinos and obtained [3,4] values similar to those 
generally accepted. In the latest issue of Particle Group Data [5], the 
following upper limit of the neutrinos mass are listed\textbf{:} $E\left( 
{\nu _{\mu} }  \right)$\textbf{ $ \le $} 0.19 MeV, \textit{E}($\nu _{\tau}  
$) $ \le $ 18.2 MeV, and \textit{E}($\nu _{e} $) $ \le $ 3 eV. In this 
contribution, based on the Expansive Nondecelerative Universe (ENU) model 
and on the ``see-saw'' mechanism [6] originally developed by Gell-Mann and 
his coworkers, the mass of all neutrinos is tentatively determined.

\bigskip

\section*{Results and discussion}

\bigskip

One of the corner stones of the ENU model states that the mass of all 
``elementary'' particles is generated by gravitational energy of the Planck 
particle (planckton) possessing the energy

\begin{equation}
\label{eq1}
E_{Pc} = m_{Pc} .c^{2} = 1.2211 \times 10^{19}GeV
\end{equation}

For the rest energy \textit{E} of any particle it thus holds [7]

\begin{equation}
\label{eq2}
E = \int {\varepsilon _{g\left( {Pc} \right)} dV_{C} \cong \frac{{m_{Pc} 
.c^{2}.\lambda _{C}} }{{a_{\left( {T} \right)}} }} 
\end{equation}

\noindent
in which $\lambda _{C} $ is the Compton wavelength, $\varepsilon _{g\left( 
{Pc} \right)} $is the planckton gravitational field density, $a_{\left( {T} 
\right)} $represents the gauge factor related to the specific time when

\begin{equation}
\label{eq3}
k.T \cong m.c^{2}
\end{equation}

In fact, equation (\ref{eq2}) expresses the gravitational energy of planckton in the 
Compton volume of a given particle. 

Neutrinos do not take part in strong and electromagnetic interactions and 
their gravitational influence can be neglected as well. The only interaction 
related to neutrinos is weak interaction mediated by vector bosons Z and W. 
Gravitational influence of the bosons on their surroundings starts just when 
their Compton wavelength becomes identical to their effective gravitational 
range [7], i.e. when

\begin{equation}
\label{eq4}
a_{Z,W} = \frac{{\hbar ^{2}}}{{G.m_{Z,W}^{3}} }
\end{equation}

\noindent
where $m_{Z,W} $is the mass of corresponding bosons. To determine a specific 
value of a neutrino mass, the value of $\lambda _{C} $ must be substituted 
by another distance closely related to the neutrino. As such a quantity, the 
effective cross section $\sigma $ of weak interactions has been selected, 
the value of which reaches

\begin{equation}
\label{eq5}
\sqrt {\sigma}  \cong 8.36 \times 10^{ - 24}m
\end{equation}

Relations (\ref{eq2}), (\ref{eq4}) and (\ref{eq5}) lead directly to the electron neutrino mass 
(energy)

\begin{equation}
\label{eq6pov}
m\left( {\nu _{e}}  \right).c^{2} = \frac{{m_{Pc} .c^{2}.\sigma 
^{1/2}}}{{a_{Z,W}} } \cong 2.2 \times 10^{ - 12}eV
\end{equation}

It was postulated by the authors of ``see-saw'' mechanism that relation of 
the masses of the neutrinos and corresponding particles (electron $e$, muon 
$\mu $ and tau-lepton $\tau $) can be formulated as follows

\begin{equation}
\label{eq6}
m\left( {\nu _{e}}  \right):m\left( {\nu _{\mu} }  \right):m\left( {\nu 
_{\tau} }  \right) = m^{2}\left( {e} \right):m^{2}\left( {\mu}  
\right):m^{2}\left( {\tau}  \right)
\end{equation}

For the electron, muon and tau-lepton their mass values have been published 
in [5] and are as follows

\begin{equation}
\label{eq7}
m\left( {e} \right) = 0.511003 MeV
\end{equation}

\begin{equation}
\label{eq8}
m\left( {\mu}  \right) = 105.658MeV
\end{equation}

\begin{equation}
\label{eq9}
m\left( {\tau}  \right) = 1.777GeV
\end{equation}

Applying (\ref{eq6pov}) to (\ref{eq9}), in addition to the electron neutrino mass, the 
following masses are obtained for $\nu _{\mu}  $ and $\nu _{\tau}  $

\begin{equation}
\label{eq10}
m\left( {\nu _{\mu} }  \right) \cong 9.4 \times 10^{ - 8}eV
\end{equation}

\begin{equation}
\label{eq11}
m\left( {\nu _{\tau} }  \right) \cong 2.7 \times 10^{ - 5}eV
\end{equation}

The results are in good agreement with those provided by GUTs within the 
SU(\ref{eq5}) approach (for the electron neutrino it gives $10^{ - 13} - 10^{ - 9}$ 
eV [8]) and they do not exceed the upper currently accepted limiting values 
[5].

\bigskip

\section*{References}

\bigskip

\noindent
1. F.J. Blatt, \textit{Modern Physics}, McGraw-Hill, New York, 1992, p. 433

\noindent
2. M. S\'{u}ken\'{\i}k, J. \v{S}ima, J. Vanko, \textit{General
  Relativity and Quantum Gravity}, Preprint gr-qc/0010061

\noindent
3. V. Skalsk\'{y}, M.S\'{u}ken\'{\i}k, \textit{Astrophys. Space Sci.,
  190} (1992) 197

\noindent
4. V. Skalsk\'{y}, M.S\'{u}ken\'{\i}k, \textit{Astrophys. Space Sci.,
  204} (1993) 161

\noindent
5. D.E. Groom et al., \textit{Eur. Phys. J. C15} (2000) 1

\noindent
6. D.W. Sciama, \textit{Nature, 348} (1990) 617

\noindent
7. J. \v{S}ima, M. S\'{u}ken\'{\i}k, \textit{General Relativity and Quantum Gravity}, 
Preprint gr-qc/0011057

\noindent
8. F. Boehm, P. Vogel, \textit{Physics of Massive Neutrinos} (2$^{nd}$ ed.), 
Cambridge University Press, 1992, p. 20

\end{document}